\documentstyle [12pt]{article}

\begin{document}
\begin{titlepage}
\begin{flushright}
\vspace*{-3cm}
hep-th/9411109  \\
TUTP-94-18  \\ November 1994
\vspace*{1cm}
\end{flushright}

\begin{center}
{\Large \bf One-Loop Renormalization of a
 Self-Interacting \\Scalar Field in Nonsimply Connected Spacetimes}\\
\vspace{.3in}{\large\em L.H.Ford and N.F.Svaiter\footnote{Permanent
address: Centro Brasileiro de Pesquisas Fisica-CBPF, Rua Dr. Xavier
Sigaud 150, Rio de Janeiro, RJ 22290-180, Brazil} }\\
 Institute of Cosmology, Department of Physics and Astronomy\\ Tufts
University, Medford, Massachusetts 02155 USA
\subsection*{\\Abstract}
\end{center}

Using the effective potential, we study the one-loop renormalization of
a massive self-interacting scalar field at finite
temperature in flat manifolds with one or more compactified
spatial dimensions.
We prove that, owing to the compactification and finite temperature,
the renormalized physical parameters of the
theory (mass and coupling constant) acquire thermal and topological
contributions. In the case of one compactified spatial dimension at finite
temperature,
we find that the corrections to
the mass are positive, but those to the coupling
constant are negative. We discuss the possibility of triviality, i.e. that
the renormalized coupling constant goes to zero at some temperature or at some
radius of the compactified spatial dimension.

\end{titlepage}
\newpage
\baselineskip 18pt
\section {Introduction}\

It is well known that one loop quantum corrections may alter the physical
parameters of an interacting quantum field theory. In general this alteration
is not of a form which can be absorbed by a simple redefinition of the
parameters, in the way that one can remove the ultraviolet divergences. A
simple example of this occurs in finite temperature field theory, where the
renormalized mass can become temperature dependent\cite{DJ}. Similarly, in flat
spacetime with compactification in one spatial direction, the mass can
depend upon the periodicity length in the compact direction\cite{FY,
Toms1,mass_gen}. This phenomenon is of particular interest in theories
with broken symmetry, as it allows both thermal and topological effects
to play a role in the breaking and restoration of symmetry\cite{sym}.
The aim of this paper is to discuss quantum field theory at finite
temperature in a spacetime where at least
one of the spatial dimensions is compactified. The particular model which
we adopt is a scalar field with quartic self-coupling. In particular, we
wish to investigate the dependence of the renormalized mass and coupling
constant upon the temperature and the size of the compactified dimension.
We will calculate the effective potential, which may be expressed in terms
of Epstein zeta functions. The ultraviolet divergences may be removed by
analytic regularization and renormalization.

The outline of this paper is the following. In Section II,
periodic boundary conditions are imposed upon
the fields (after a Wick rotation), and the temperature dependent one-loop
effective potential is calculated.
The theory is regularized using an analytic
continuation of the inhomogeneous Epstein zeta function. The renormalization
of $\lambda\varphi^{4}$ theory in this multiply
connected spacetime can be done
by introducing counterterms, and we show that the mass and coupling constant
counterterms are temperature and size independent

In Section III, we assume finite temperature and only one compactified
spatial dimension. We explicitly calculate the corrections to both the mass
and the coupling constant in this case. We find that the corrections
to the mass are positive but those to the coupling constant are negative.
The results are discussed in Section IV. In particular, we discuss the
possibility of arranging for the renormalized coupling constant to vanish
(``triviality'') at some particular temperature or spatial size.
In this paper we use units in which $\hbar=c=k_{B}=1$.

\section{The effective potential of a scalar field at finite
temperature}\

In this section we study a real massive scalar field at finite temperature,
where we assume that
the topology of the spacelike sections is that of a three-torus.
This kind of topology
allows two different types of scalar fields. One which is periodic in the
identified spatial coordinates is called an untwisted field, and the other
which is antiperiodic in the identified spatial coordinate is called a twisted
scalar field\cite{Isham}.
To study twisted scalar fields, we cannot assume that the normalized vacuum
expectation value of the field is constant and the
effective potential cannot be used. For the sake of simplicity, in
this paper we will study only the untwisted scalar field.

The Lagrange density of the field is
\begin{equation}
{\cal L}= \frac{1}{2}\partial_{\mu}\varphi_{u}\partial^{\mu}\varphi_{u}
-\frac{1}{2}m^{2}_{0}~
\varphi_{u}^{2}-\frac{\lambda_{0}}{4!}\varphi^{4}_{u}\,
\end{equation}
where $\varphi_{u}(x)$ is the unrenormalized field,
and $m_{0}$ and $\lambda_{0}$
are the bare mass and coupling constant, respectively. We may rewrite the
Lagrange density in the usual form where the counterterms will appear
explicitly. Defining
\begin{equation}
\varphi_{u}(x)= (1+\delta Z)^{\frac{1}{2}}\varphi(x)
\end{equation}
\begin{equation}
m^{2}_{0}=(m^{2}+\delta m^{2}) (1+\delta Z )^{-1}
\end{equation}
\begin{equation}
\lambda_{0}= (\lambda+\delta\lambda)(1+\delta Z)^{-2}
\end{equation}
and substituting Eqs. (2), (3) and (4) into Eq. (1), we have
\begin{equation}
{\cal L}=\frac{1}{2}\partial_{\mu}\varphi\partial^{\mu}\varphi
-\frac{1}{2}m^{2}\varphi^{2}-
\frac{\lambda}{4!}\varphi^{4}+\frac{1}{2}\delta Z~\partial_{\mu}\varphi
\partial^{\mu}\varphi-
\frac{1}{2}\delta m^{2}\varphi^{2}-\frac{1}{4!}\delta\lambda~\varphi^{4}\,
\end{equation}
where $\delta Z$, $\delta m^{2}$, and $\delta\lambda$ are the wave function,
mass and coupling constant counterterms of the model. Through this paper
we will assume that $m^{2}>0$. In the one-loop
approximation, the effective potential at zero temperature in
uncompactified spacetime is given by \cite{CW,Jackiw}
\begin{eqnarray}
V(\varphi_{0})&=& \frac{1}{2}m^{2}\varphi^{2}_{0}+\frac{\lambda}{4!}
\varphi^{4}_{0}-\frac{1}{2}\delta m^{2}\varphi^{2}_{0}-
\frac{1}{4!}\delta\lambda~\varphi^{4}_{0}\  \nonumber \\&+&i\int d^{4}q
\frac{1}{(2\pi)^{4}}
\sum_{s=1}^{\infty}\frac{1}{2s}\biggl(\frac{1}{2}\lambda
\varphi^{2}_{0}\biggr)^{s}\frac{1}{(q^{2}-m^{2}+i\epsilon )^s}\, .
                                       \label{eq:V}
\end{eqnarray}

There is no difficulty in extending the above results to finite temperature
states. In this case, functional integrals will run over the fields that
satisfy periodic boundary conditions in Euclidian time. The effective action
can be defined as in the zero temperature case by a functional Legendre
transformation, and regularization and renormalization procedures follow
the same steps as in the zero temperature case. Similarly, compactification
in imposed by requiring that the field be periodic in the spatial directions.

It has been shown that for models where the spacelike section are noncompact,
all the divergences present in the Feynman loops are temperature
independent\cite{KM,MOU}. Similarly, the renormalization of the zero
temperature
theory with at least one compactified spatial dimension has been investigated
by Toms\cite{Toms2} and by Birrell and Ford\cite{BF}.
These authors found that through the two-loop level, all of the counterterms
are independent of the spatial size. A more general discussion has been
given by Banach\cite{Banach}, who shows that topological identifications
will not introduce new counterterms.
Thus the divergences of the theory are independent of both
temperature and spatial size. If this were not the case, there would be
a danger that the renormalizability of the theory would be upset by
changing either the temperature or the spatial topology.

Let us assume that we have a massive scalar field at finite temperature
$ \beta^{-1}$, and that the spacelike section is compactified
with the topology of a three torus of sides
$L_{1},L_{2}$ and $L_{3}$. Define
\begin{equation}
c^{2}=\frac{m^{2}}{4\pi^{2}\mu^{2}}
\end{equation}
\begin{equation}
(\beta\mu)^{2}=a^{-1}_{4}
\end{equation}
\begin{equation}
(L_{i}\mu)^{2}=a^{-1}_{i} ~~~ i=1,2,3 \quad ,
\end{equation}
where $\mu$ is a mass parameter introduced to keep the Epstein zeta function
a dimensionless quantity. The euclidean effective potential becomes
\begin{eqnarray}
V_{E}(\beta,L_{1},L_{2},L_{3},\varphi_{0})&=&\frac{1}{2}m^{2}\varphi^{2}_{0}
+\frac{\lambda}{4!}
\varphi^{4}_{0}-\frac{1}{2}\delta m^{2}\varphi^{2}_{0}-\frac{1}{4!}
\delta\lambda\varphi^{4}_{0} \nonumber \\&+&\frac{1}{\beta L_{1}L_{2}L_{3}}
\sum^{\infty}_{s=1}\frac{(-1)^
{s+1}}{2s}\biggl(\frac{\lambda}{8\pi^{2}}\biggr)^{s}
\biggl(\frac{\varphi_{0}}{\mu}\biggr)^{2s}A^{c^{2}}_{4}
(s,a_{1},a_{2},a_{3},a_{4}) \, , \label{eq:VE}
\end{eqnarray}
where
\begin{equation}
A^{c^{2}}_{N}(s,a_{1},a_{2},..,a_{N})=\sum^{\infty}_{n_{1},n_{2}..n_{N}=-\infty}
(a_{1}n_{1}^{2}+a_{2}n^{2}_{2}+...+a_{N}n^{2}_{N}+c^{2})^{-s}
\end{equation}
is the inhomogeneous Epstein zeta function\cite{Epstein, Erdelyi}.
Note that in going from Eq.~(\ref{eq:V}) to Eq.~(\ref{eq:VE}), we have
first performed a Wick rotation so that the momenta are euclidean,
and then have replaced the momentum integrals by discrete sums.
In the case $c^{2}=0$,
Eq. (11) defines a Madelung sum in the theory of classical lattices. If we
impose the condition that the renormalized mass is zero, there is a
problem in defining the renormalized coupling constant. The way to
circumvented this difficulty is to impose the renormalizations conditions
not at $\varphi_{0}=0$ but at another point. For a careful
discussion, see the paper by Coleman and Weinberg\cite{CW}.
In this paper we assume $m^{2}>0$, and the above problem does not
appear.
In the limit $ L_{i}\rightarrow\infty $,
the expression given by Eq. (10) differs from the usual finite temperature
effective potential by terms that
are independent of $ \varphi_{0}$ \cite{Linde}. Because only derivatives
with respect to $\varphi_{0}$ correspond to
physically meaningful quantities, this does not pose any problems.

Let us define the modified inhomogeneous Epstein zeta function as
\begin{equation}
E^{c^{2}}_{N}(s,a_{1},a_{2},..a_{N})=\sum^{\infty}_{n_{1},n_{2},..n_{N}=1}
(a_{1}n^{2}_{1}+..+a_{N}n^{2}_{N}+c^{2})^{-s}
\end{equation}
A simple calculation gives
\begin{eqnarray}
A^{c^{2}}_{4}(s,a_{1},a_{2},a_{3},a_{4})&=&
16E^{c^{2}}_{4}(s,a_{1},a_{2},a_{3},a_{4})+8 E^{c^{2}}_{3}
(s,a_{1},a_{2},a_{4})+8 E^{c^{2}}_{3}(s,a_{1},a_{3},a_{4}) \nonumber \\&+&8
 E^{c^{2}}_{3}(s,a_{2},a_{3},a_{4})+
8 E^{c^{2}}_{3}(s,a_{1},a_{2},a_{3})+
4 E^{c^{2}}_{2}(s,a_{1},a_{2}) \nonumber \\&+&4
E^{c^{2}}_{2}(s,a_{1},a_{3})+4 E^{c^{2}}_{2}(s,a_{1},a_{4})+
4 E^{c^{2}}_{2}(s,a_{2},a_{3}) \nonumber \\&+&4
E^{c^{2}}_{2}(s,a_{2},a_{4})+ 4 E^{c^{2}}_{2}(s,a_{3},a_{4})
+ 2 E^{c^{2}}_{1}(s,a_{1}) \nonumber \\&+&2
E^{c^{2}}_{1}(s,a_{2})
+ 2 E^{c^{2}}_{1}(s,a_{3})+ 2 E^{c^{2}}_{1}(s,a_{4})+
c^{-2s}.
\end{eqnarray}
Defining
the new coupling constant and a dimensionless vacuum expectation value of the
field by
\begin{equation}
g=\frac{\lambda}{8\pi^{2}}
\end{equation}
\begin{equation}
\frac{\varphi_{0}}{\mu}=\phi \, ,
\end{equation}
the finite temperature one-loop effective potential is given by
\begin{eqnarray}
V_{E}(\beta,L_{1},L_{2},L_{3},\phi)&=& \mu^{4}\biggl(2\pi^{2}
c^{2}\phi^{2}
+\frac{1}{3}\pi^{2} g \phi^{4}
-\frac{1}{2\mu^{2}}{\delta m^{2}}\phi^{2}-\frac{1}{4!}
\delta\lambda\phi^{4}\biggr) \nonumber \\&+&
\frac{1}{\beta L_{1}L_{2}L_{3}}\sum^{\infty}_{s=1}
\frac{(-1)^{s+1}}{2s}
g^{s}\phi^{2s}A^{c^{2}}_{4}(s,a_{1},a_{2},a_{3},a_{4}) \, .
\end{eqnarray}

It is possible to regularize the one-loop effective potential introducing
a cutoff in the Euclidian region, but we prefer to use
the method
of analytic extension.
Let us assume that each term of the series in $s$ in
the one-loop effective potential $ V_{E}(\beta,L_{1},L_{2},L_{3}
,\varphi_{0})$ is an analytic extension,
defined in the beginning only in an open connected set.
To render the discussion more general, let us discuss the
process of the analytic continuation of the modified inhomogeneous Epstein
zeta function given by Eq. (12). For $ Re(s) > \frac{N}{2}$, the $E^{c^{2}}
_{N}(s,a_{1},a_{2},..a_{N}) $ converges and represent an analytic function
of $ s$, so $s > \frac{N}{2} $ is the largest possible domain of the
convergence of the series. This means that in Eq. (10) only the terms
$ s=1 $ and $ s=2$ are divergent. The $s=1$ term arises from the self-energy
diagram (the one-loop process with two external lines), and the $s=2$
term arises from the one-loop correction to the scattering amplitude
(the one-loop diagram with four external lines).
After regularization, we may think of the first two terms in the sum in
Eq.~(\ref{eq:VE}) as being evaluated not at $s=1$ and $s=2$, but rather
at $s=1+\alpha$ and $s=2+\alpha$, respectively, where $\alpha$ is a
complex parameter which vanishes in the limit in which the regularization
is removed.

Using a Melin transform, it is possible to continue analytically $ E^{c^{2}}
_{N}(s,a_{1},..,a_{N})$ from $ Re(s) >
\frac{N}{2} $ to $ Re(s) \leq \frac{N}{2}$,
although isolated singularities will appear in the closed region
$ Re(s) \leq \frac{N}
{2}$ at the points
\begin{equation}
s=\frac{N}{2},\, \frac{N-1}{2},\, \cdots \, \frac{1}{2},\,
                                       -\frac{2l+1}{2},~~l \in N.
\end{equation}
At these points, the analytic extension of $ E^{c^{2}}_{N}(s,a_{1}..,a_{N}) $
has first order poles, with residues $ Res\biggl[E^{c^{2}}_{N}(s,a_{1},..,a_
{N}),s_{i}\biggr]$. The exact
expression of the residue at the points in which we are interested is
\begin{equation}
Res \biggl[E^{c{2}}_{N}(s,a_{1},...,a_{N}),\frac{j}{2}\biggr]=
\frac{(-1)^{N-j}\pi^{\frac{j}{2}}}
{2^{N}\Gamma(\frac{j}{2})}\sum^{\frac{N-j}{2}}
_{k=0}\frac{(-1)^{k}}{k!}c^{2k}\pi^{k} A(2k+j)\,
\end{equation}
where
\begin{equation}
A(k)=\sum_{\{i_{1},..,i_{k}\}}\sqrt{a_{i_{1}}...a_{i_{k}}}\
,
\end{equation}
and $\sum_{i_{1},..i_{k}} $ denotes the sum over all possible choices of
the $ i_{1},..i_{k}$ among $1,...N$ (for $k=0$ the sum is set equal to one)
\cite{Kirstein}. An appropriate choice of $\delta m^{2}$ and $\delta\lambda$
will remove the poles at $s=1$ and $s=2$, respectively.
The idea to continue analytically
expressions and subtract the poles was
exploited by various authors\cite{BGD,Speer}.
In the method used by Bollini, Giambiagi and
Domingues\cite{BGD}, a complex parameter
was introduced as an exponent of the denominator of
the loop expressions and
the integrals are well defined analytic functions of the parameter for
some $ s_{0}$. Performing an analytic extension of this expression for
$ s < s_{0} $, poles will appear in the analytic extension and the final
expression becomes finite after the subtraction
of the poles. It is clear that our regularization is exactly
a discrete version of the Bollini et al analytic regularization.

In our problem, the renormalization conditions are given by
\begin{equation}
\frac{\partial^{2}}{\partial\phi^{2}}V_{E}(\beta,L_{1},L_{2},L_{3},\phi)
|_{\phi=0}=
4\pi^{2}\mu^{4}c^{2} = \mu^2 m^2
\end{equation} and
\begin{equation}
\frac{\partial^{4}}{\partial\phi^{4}}V_{E}(\beta,L_{1},L_{2},L_{3},\phi)
|_{\phi=0}=8\pi^{2}\mu^{4}g = \mu^4 \lambda \, .
\end{equation}
Substituting Eq. (16) into Eqs. (20) and (21) in
such a way that the counterterm
$\delta m^{2}$ cancels the pole contribution at $s=1$ of the analytic
extension of the inhomogeneous Epstein zeta function, and $\delta\lambda$
cancels the pole contribution at $s=2$, we have
\begin{equation}
\delta m^{2}=\frac{g}{\beta L_{1}L_{2}L_{3}\mu^{2}}
\frac{1}{s-1}
Res\biggl[A^{c^{2}}_{4}(s,a_{1},a_{2},a_{3},a_{4}),s=1\biggr]
\end{equation}
and
\begin{equation}
\delta\lambda=-24\frac{g^{2}}{\beta L_{1}L_{2}L_{3}\mu^{4}}
\frac{1}
{s-2} Res\biggl[A^{c^{2}}_{4}(s,
a_{1},a_{2},a_{3},a_{4}),s=2\biggr].
\end{equation}
By substitution of Eq. (18) and Eq. (19) into Eq. (22) and Eq. (23), it
is straightforward to
show that both $\delta\lambda$ and $\delta m^{2}$ are temperature and size
independent. This shows that in the one-loop approximation,
the counterterms of the model are independent
of the parameters which are associated with the nontrivial topology. Hence
if the model is renormalizable at zero temperature with certain
counterterms, it is also renormalizable at finite temperature with exactly
the same counterterms.

\section{Topological and Thermal Corrections to the Mass and Coupling
Constant}

In this section we will investigate  the
thermal and topological correction to the renormalized mass and
coupling constant in the case where there is compactification in only
one spatial direction. Set $L=L_3$ and take the limit in which
$L_1 \rightarrow \infty$ and $L_2 \rightarrow \infty$.
The finite temperature one-loop
effective potential in this case is given by:
\begin{equation}
V_{E}(\beta,L,\phi)=\mu^{4}\biggl(2\pi^{2}c^{2}\phi^{2}+
\frac{1}{3}\pi^{2}g
\phi^{4}-\frac{1}{2\mu^{2}}\delta m^{2}\phi^{2}-
\frac{1}{4!}\delta\lambda\phi^{4}\biggr)+
G_{E}(\beta,L,\phi)
\end{equation}
where
\begin{equation}
G_{E}(\beta,L,\phi)=\mu^{4}\sqrt{a_{3}a_{4}}\sum^{\infty}_{s=1}
\frac{(-1)^{s+1}}{2s}g^{s}\phi^{2s}\int d^{2}k A^{M^{2}}_{2}(s,a_{3},a_{4}),
\end{equation}
In the above equation
$$
M^{2}=(k^{1})^{2}+(k^{2})^{2}+c^{2},
$$ where
$$ k^{1}=\frac{q^{1}}{2\pi\mu}$$ and
$$ k^{2}=\frac{q^{2}}{2\pi\mu}$$
are dimensionless quantities.

Using the identity
\begin{equation}
\int \frac{d^{d}l}{(l^{2}+a^{2})^{b}}=
\frac{\pi^{d/2}}{\Gamma(b)}\Gamma(b-\frac{d}{2})a^{d-2b},
\end{equation}
we can perform the integrations over the continuous momenta and write
\begin{equation}
G_{E}(a_{3},a_{4},\phi)=
\mu^{4}\sqrt{a_{3}a_{4}}\pi\sum^{\infty}_{s=0}\frac{(-1)^{s+2}}{2s+2}
g^{s+1}\phi^{2s+2}
\frac{\Gamma(s)}{\Gamma(s+1)}A^{c^{2}}_{2}(s,a_{3},a_{4})\, .
                                                     \label{eq:GE}
\end{equation}
Note that the poles which were at $s=1$ and $s=2$ in Eq.~(\ref{eq:VE})
are now located at $s=0$ and $s=1$, respectively, in Eq.~(\ref{eq:GE}).
As before, we regularize Eq.~(\ref{eq:GE}) by analytically continuing
the summand around the points $s=0$ and $s=1$.

    The formal correction to the squared mass is given by
\begin{equation}
\Delta' m^2 = \left(\frac{d^2 G_E}{d \varphi_0^2} \right)_{s=0}
= \pi g \mu^2\, \sqrt{a_3 a_4}\; \lim_{s\rightarrow 0}\; \bigl[ \Gamma(s)
A_2^{c^2}(s,a_3,a_4) \bigr] \,.  \label{eq:formasscorr}
\end{equation}
The $s\rightarrow 0$ limit of the Epstein zeta function is evaluated in the
Appendix. The result, Eq.~(\ref{eq:A2rep2a}), may be used to separate
$\Delta' m^2$ into its infinite and finite parts:
\begin{equation}
\Delta' m^2 = \delta m^2 + \Delta m^2 \,,
\end{equation}
where $\delta m^2$ is the counterterm to be absorbed by mass renormalization,
and $\Delta m^2$ is the finite correction to the mass. The latter is defined
so as to vanish in the limit of zero temperature in noncompactified space.
Explicitly, we have
\begin{equation}
\delta m^2 = g \mu^2 \biggl(-\frac{\pi c^2}{s} + 2\pi \ln c \biggr)\,,
\end{equation}
and
\begin{equation}
\Delta m^2 = \frac{\lambda}{4\pi^2} \Biggl[ m^2 \int_1^\infty
{{(t^2-1)^{{1\over 2}} dt} \over {{\rm e}^{m\beta t} -1}} \,
- \frac{\pi}{\beta L} \sum_{n =-\infty}^{\infty}
\ln \Bigl(1 - e^{-2\pi L \beta^{-1} \sqrt{n^2 +m^2
\beta^2/4\pi^2}}\Bigr)\Biggr]
   \, .   \label{eq:masscorr}
\end{equation}
There is an equivalent expression obtained by interchange of $\beta$ and $L$.
Note that $\Delta m^2 \geq 0$ for all choices of the parameters.
The first term in Eq.~(\ref{eq:masscorr}) is the purely thermal
correction to the mass\cite{DJ}. In the limit that
$L \rightarrow \infty$, the second
term vanishes, and this correction is all that survives. The second term
becomes the purely topological correction\cite{Toms1}
in the limit of zero temperature:
\begin{equation}
\Delta m^2 \sim -\frac{\lambda}{2\pi L}\, \int_0^\infty dx \,
\ln \Bigl(1 - e^{-2\pi L \sqrt{x^2 +m^2/4\pi^2}}\Bigr) =
\frac{\lambda}{4\pi^2}  m^2 \int_1^\infty
{{(t^2-1)^{{1\over 2}} dt} \over {{\rm e}^{mL t} -1}} \, .
\end{equation}
(An integration by parts was performed to obtain the last form.)

The formal correction to the coupling constant $\lambda$ due to finite
temperature and/or spatial compactification is given by
\begin{equation}
\Delta' \lambda = \left(\frac{d^4 G_E}{d \varphi_0^4} \right)_{s=1}
 = -\frac{3\lambda^2}{32\pi^3}\, \sqrt{a_3 a_4} \,
                  \lim_{s\rightarrow 1} A_2^{c^2}(s,a_3,a_4) \,.
\end{equation}
This quantity is, of course, ill-defined because of a pole term at $s=1$
which needs to be isolated and removed. In the Appendix, it is shown that
\begin{equation}
A_2^{c^2}(s,a_3,a_4) \sim
{\pi \over {\sqrt{a_3 a_4}}} \left[ {1 \over {s-1}} - \ln c^2 + \cdots \right]
+ F_1(a_3,a_4) \, , \qquad s \rightarrow 1 \,.
\end{equation}
The pole term is absorbed by the $\delta \lambda$ counterterm when we let
\begin{equation}
\delta \lambda =  -\frac{3\lambda^2}{32\pi^2}
\left( {1 \over {s-1}} - \ln c^2  \right) \,.
\end{equation}
Then the finite correction to the coupling constant is given by
\begin{equation}
\Delta \lambda =  -\frac{3\lambda^2}{32\pi^3}\, \sqrt{a_3 a_4}\,
                                  F_1(a_3,a_4) \,. \label{eq:delcc1}
\end{equation}
Here
\begin{equation}
F_1(a_3,a_4) =\frac{1}{{\sqrt{a_3 a_4}}}\left[f(a_3) + f(a_4) +
                    R(a_3,a_4) \right] \,,  \label{eq:F1}
\end{equation}
where
\begin{equation}
f(a_3) = 4\pi \int_1^\infty {{(t^2-1)^{-\frac{1}{2}} dt}
                         \over {{\rm e}^{2\pi c t/ \sqrt{a_3}} -1}} \,,
\end{equation}
and $R(a_3,a_4)$ is given by
\begin{eqnarray}
R(a_3,a_4) &=& 2\pi \sqrt{a_3} \sum^\infty_{n=-\infty}
\frac{1}{\sqrt{a_3 n^2+ c^2}\left(e^{2\pi\sqrt{(a_3 n^2+ c^2)/a_4}}-1\right)}
-f(a_4) \nonumber \\
  &=& 2\pi \sqrt{a_4} \sum^\infty_{n=-\infty}
\frac{1}{\sqrt{a_4 n^2+ c^2}\left(e^{2\pi\sqrt{(a_4 n^2+ c^2)/a_3}}-1\right)}
-f(a_3) \,. \label{eq:Rrep}
\end{eqnarray}

    Note that the thermal and topological corrections to the coupling constant
are always negative:
\begin{equation}
\Delta \lambda < 0 \,,
\end{equation}
which follows, for example, from Eqs.~(\ref{eq:F1}-\ref{eq:Rrep}),
where it is apparent that
${F_1}(a_3,a_4) > 0$. The three terms on the right-hand side of
Eq.~(\ref{eq:delcc1}) can each be given a physical interpretation. The
$f(a_3)$ term is the purely topological term. It is the correction to the
coupling constant at zero temperature in a space with one compact dimension.
Similarly, the $f(a_4)$ term is the purely thermal term, which is the
correction to the coupling constant at finite temperature in uncompactified
space. The $R$ term represents a coupling between thermal and topological
effects which is present only at finite temperature in compactified space.

    It is of interest to examine the small mass limit of this correction
to the coupling constant. In the limit that $m \rightarrow 0$, the dominant
contribution to $\Delta \lambda$ comes from the $n=0$ term in $R(a_3,a_4)$,
and we obtain
\begin{equation}
\Delta \lambda \sim  -\frac{3\lambda^2}{8m^2 \beta L}, \qquad m \rightarrow 0.
\label{eq:smallm}
\end{equation}
This result seems to indicate that the one loop correction to the coupling
constant can be arbitrarily negative for small masses. However, one must
be careful about the limits of validity of the one loop approximation.
This issue will be discussed in more detail in the next section.
Note that the coupling
constant correction described by Eq.~(\ref{eq:smallm}) is nontrivial
only at finite temperature in compactified spacetime, i.e. when both
$L$ and $\beta$ are finite.

    Let us now consider the purely thermal correction in more detail. Let
$L \rightarrow \infty$, so that $a_3 \rightarrow 0$. Then
\begin{equation}
\Delta \lambda =   -\frac{3\lambda^2}{32\pi^3} f(a_4)
               =  -\frac{3\lambda^2}{8\pi^2}\,
\int_1^\infty {{(t^2-1)^{-\frac{1}{2}} dt}
                         \over {{\rm e}^{\beta m t} -1}} \,.
                                 \label{eq:delcc2}
\end{equation}
In the low temperature limit($\beta \rightarrow \infty$), we have
\begin{equation}
\Delta \lambda \approx  -\frac{3\lambda^2}{8\pi^2}\,
\int_1^\infty (t^2-1)^{-\frac{1}{2}} {\rm e}^{- \beta m t} dt
= -\frac{3\lambda^2}{8\pi^2} K_0(\beta m) \approx
-\frac{3\lambda^2}{8\pi^2} \sqrt{\frac{\pi}{2\beta m}}\, {\rm e}^{-\beta m} \,.
                                 \label{eq:delcc3}
\end{equation}
Similarly, in the high temperature limit ($\beta \rightarrow 0$), we
may use
\begin{equation}
\frac{1}{{\rm e}^{2\pi \beta m t} -1} \approx \frac{1}{\beta m t}
                       = \frac{T}{m t}
\end{equation}
to write
\begin{equation}
\Delta \lambda \sim -\frac{3\lambda^2 T}{16\pi m}\,,
                     \qquad T \rightarrow  \infty \,. \label{eq:highT}
\end{equation}
Again, this correction would seem to be large in the case that $T \gg m$.

   In this section we have found that in a space with one compact spatial
dimension and/or at finite temperature, the one-loop correction to the
squared mass is always positive, whereas that to the coupling constant
is always negative.  In the limits that either the temperature vanishes,
or that the size of the compact dimension becomes large, we recover
the results of previous authors for the mass correction, $\Delta m^2$.
To our knowledge, the results presented here for the coupling constant
correction, $\Delta \lambda$, have not been given before. The only
reference of which we are aware which discusses either thermal or
topological corrections to coupling constants is Higuchi and
Parker\cite{HP}. Of course, all of the results of this section also
apply to the case of a spacetime with periodicity in two spatial directions,
but at zero temperature. One simply replaces $L$ and $\beta$ in the above
formulas by $L_1$ and $L_2$, the two periodicity lengths.

\section{Discussion}\

      In this paper, we have calculated the mass correction, $\Delta m^2$,
and the coupling constant correction, $\Delta \lambda$, due to both finite
temperature and compactification in one spatial direction. We found that
$\Delta m^2 \ge 0$, whereas $\Delta \lambda \le 0$. One of the primary
reasons for interest in $\Delta m^2 $ is its role in symmetry restoration.
It had been noted by previous authors that  $\Delta m^2 \ge 0$ when one
has either finite temperature in uncompactified space, or compactification
at zero temperature. Thus in both cases, the effect of the radiative
correction is to restore broken symmetries. Our results show that this effect
also holds when one has a finite temperature state in compactified space.
The extension of these results to models in spacetime dimensions other
than four and to other model field theories is undertaken in a separate
paper\cite{MS}.

     An interesting feature of the negative coupling constant correction
is that it tends to make the theory less strongly coupled. One is then
tempted to raise the question of whether it would even be possible to
cause the net coupling constant to vanish, i.e., to achieve triviality
at some particular temperature or compactification length. We have defined
$\lambda$ to denote the renormalized coupling constant at zero temperature
in uncompactified space. Thus the effective coupling constant when either
$L$ or $\beta$ are finite is
\begin{equation}
\lambda' =  \lambda + \Delta \lambda \, .
\end{equation}
The one-loop correction,  $\Delta \lambda$, is of order $\lambda^2$, so it
is not clear that one can make it equal to $\lambda$ in magnitude before
the one-loop  approximation fails. The crucial issue here is just what are
the limits of validity of this approximation. If it is simply that one
needs $\lambda \ll 1$, then this does not prevent us from arranging a
situation where $\lambda' =0$. This is apparent from Eqs.~(\ref{eq:smallm})
or (\ref{eq:highT}). However, it is not clear that the true expansion
parameter is $\lambda$ itself. It could  be $\lambda$ multiplied
by a dimensionless function of $m$, $L$, and $T$.  If, for example, the
limit of validity of Eq.~(\ref{eq:highT}) is when $\lambda T/m \ll 1$,
then it is not possible to use this relation at the point where $\lambda'$
would vanish. To settle this question, it would be necessary to have a
reliable estimate of the magnitude of the higher order corrections.

\section{Acknowledgement}

We would like to thank Prof. A. Vilenkin for several helpful
discussions. N.F. Svaiter would like to  acknowledge the
hospitality of the Institute of Cosmology, Tufts University, where part of this
work was carried out . This work was supported by Conselho Nacional de
Desenvolvimento Cientifico e Tecnologico do Brazil (CNPq) and by National
Science Foundation Grant PHY-9208805.

\appendix
\section*{Appendix}
\setcounter{equation}{0}
\renewcommand{\theequation}{A\arabic{equation}}

   In this appendix, we wish to derive an expression for the Epstein zeta
function $A_2^{c^2}(s,a_1,a_2)$. This quantity is initially defined by the
double sum
\begin{equation}
A_2^{c^2}(s,a_1,a_2) =
\sum_{n_1, n_2 =-\infty}^\infty \bigl(a_1 n_1^2 +a_2 n_2^2 +c^2 \bigr)^{-s}
\, . \label{eq:A2def}
\end{equation}
This series is convergent for ${\rm Re}\, s > 2$, and divergent otherwise.
Our starting point will be the following summation formula, which is proven
in Ref. \cite{F80}  :
\begin{equation}
F(\lambda,a) = \sum_{n =-\infty}^\infty \bigl( n^2 +a^2 \bigr)^{-\lambda}
= a^{1-2\lambda} \Biggl[ \sqrt{\pi}\, {{\Gamma(\lambda-{1\over 2})} \over
{\Gamma(\lambda)}}
+ 4 \sin{\pi\lambda} \int_1^\infty
{{(t^2-1)^{-\lambda} dt} \over {{\rm e}^{2\pi a t} -1}} \Biggr] \, .
                                       \label{eq:sumfor}
\end{equation}
The series representation of $F$ converges for
${\rm Re}\, \lambda > {1\over 2}$,
whereas the integral representation is defined for ${\rm Re}\, \lambda < 1$.

   Let us first use Eq.~(\ref{eq:sumfor}) to replace the $n_1$ summation in
Eq.~(\ref{eq:A2def}). The result is
\begin{eqnarray}
A_2^{c^2}(s,a_1,a_2) &=&
                  \sqrt{\pi}\, {{\Gamma(s-{1\over 2})} \over{\Gamma(s)}}
a_1^{-{1\over 2}} a_2^{{1\over 2}-s}\, F(s -{1\over 2},{c\over{\sqrt{a_2}}})
\nonumber \\
&+& 4 a_1^{-s} \sin{\pi s}\, \sum_{n =-\infty}^\infty a^{1-2s} \int_1^\infty
{{(t^2-1)^{-s} dt} \over {{\rm e}^{2\pi a t} -1}} \, , \label{eq:A2rep1}
\end{eqnarray}
where
\begin{equation}
a = \sqrt{\frac{a_2 n^2 +c^2}{a_1}}\, . \label{eq:defa}
\end{equation}
The function $F(s -{1\over 2})$ which
appears in the first term of Eq.~(\ref{eq:A2rep1}) can in turn be expressed
as an integral using Eq.~(\ref{eq:sumfor}) a second time. The result is
\begin{eqnarray}
\sqrt{a_1 a_2} \, A_2^{c^2}(s,a_1,a_2) &=&
\pi\, {{\Gamma(s-1)} \over{\Gamma(s)}}\, c^{2(1-s)}
+\, \sqrt{\pi}\, {{\Gamma(s-{1\over 2})} \over{\Gamma(s)}}\,
\sin{\pi(s-{1\over 2})}\, c^{2(1-s)}\, \int_1^\infty
{{(t^2-1)^{{1\over 2}-s} dt} \over {{\rm e}^{2\pi c t/\sqrt{a_2}} -1}} \,
 \nonumber \\
&+& 4 a_1^{{1\over 2}-s} a_2^{{1\over 2}}\, \sin{\pi s}\,
\sum_{n =-\infty}^\infty a^{1-2s} \int_1^\infty
{{(t^2-1)^{-s} dt} \over {{\rm e}^{2\pi a t} -1}} \, , \label{eq:A2rep2}
\end{eqnarray}

    For the purpose of explicitly calculating the mass correction,
Eq.~(\ref{eq:formasscorr}),
we need to evaluate the function $\Gamma(s)\,A_2^{c^2}(s,a_1,a_2)$ in the
neighborhood of $s=0$. The pole term will arise from the factor of
$\Gamma(s-1)$ in the first term of Eq.~(\ref{eq:A2rep2}). The remaining two
terms will be finite in the $s \rightarrow 0$ limit. The required asymptotic
form is
\begin{eqnarray}
\sqrt{a_1 a_2} \,\Gamma(s)\, A_2^{c^2}(s,a_1,a_2) \sim
-\frac{\pi c^2}{s} + 2\pi \ln c
+\, 8\pi c^2\, \int_1^\infty
{{(t^2-1)^{{1\over 2}} dt} \over {{\rm e}^{2\pi c t/\sqrt{a_2}} -1}} \,
 \nonumber \\
+\, 4 \sqrt{a_1 a_2}\, \pi \,
\sum_{n =-\infty}^\infty a^{1-2s} \int_1^\infty
{{dt} \over {{\rm e}^{2\pi a t} -1}}\, , \qquad s \rightarrow 0 \, .
                                                 \label{eq:A2rep2a}
\end{eqnarray}

   For the purpose of calculating the corrections to the coupling constant,
we need to examine the $s \rightarrow 1$ limit of Eq.~(\ref{eq:A2rep2}).
There will again be a simple pole coming from the factor
of $\Gamma(s-1)$, and all other terms will be regular.
However, the $s \rightarrow 1$ limit of the third term on the right
hand side of Eq.~(\ref{eq:A2rep2})
requires some care. There is a pole coming from the integral which is cancelled
by the $\sin{\pi s}$ factor. To calculate this explicitly, let
\begin{equation}
I(s) \equiv \int_1^\infty {{(t^2-1)^{-s} dt} \over {{\rm e}^{2\pi a t} -1}}
= \int_1^{t_0} {{(t^2-1)^{-s} dt} \over {{\rm e}^{2\pi a t} -1}} +
\int_{t_0}^\infty {{(t^2-1)^{-s} dt} \over {{\rm e}^{2\pi a t} -1}} \, ,
\end{equation}
where $0<t_0 -1 \ll 1$. As $s \rightarrow 1$, the contribution of the second
integral is finite, so we may write
\begin{equation}
I(s) \sim
\int_1^{t_0} {{(t-1)^{-s} (t+1)^{-s} dt} \over {{\rm e}^{2\pi a t} -1}} \sim
{1\over {2({\rm e}^{2\pi a } -1})} \int_1^{t_0} (t-1)^{-s} dt \, .
\end{equation}
Furthermore, because $s<1$,
\begin{equation}
\int_1^{t_0} (t-1)^{-s} dt = {{(t_0 -1)^{1-s}} \over {1-s}} \sim {1 \over
{1-s}}
   \, .
\end{equation}
Next, we may use $\sin{\pi s} \sim \pi(1-s)$ as $s \rightarrow 1$ to write
\begin{equation}
\sin{\pi s}\,I(s) \rightarrow {\pi \over {2({\rm e}^{2\pi a} -1})}\, ,
 \qquad s \rightarrow 1 \, .
\end{equation}
Finally, we may combine the above results to yield an expression for
$A_2^{c^2}(s,a_1,a_2)$ in the $s \rightarrow 1$ limit:
\begin{eqnarray}
A_2^{c^2}(s,a_1,a_2) &\sim&
{\pi \over {\sqrt{a_1 a_2}}} \left[ {1 \over {s-1}} - \ln c^2 + \cdots \right]
 \nonumber \\
&+& {4\pi \over {\sqrt{a_1 a_2}}}
\int_1^\infty {{(t^2-1)^{-\frac{1}{2}} dt}
                         \over {{\rm e}^{2\pi c t/ \sqrt{a_2}} -1}}
+ {2\pi \over {a_1}} \sum_{n =-\infty}^\infty
  \frac{1}{a({\rm e}^{2\pi a} -1)} \, , \label{eq:A2rep3}
\end{eqnarray}
where $a$ is defined by Eq.~(\ref{eq:defa}).

     Note that $A_2^{c^2}(s,a_1,a_2) =A_2^{c^2}(s,a_2,a_1)$, whereas the
procedure which we have used has obscured this symmetry. Thus, there is
an alternative expression for $A_2^{c^2}(s,a_1,a_2)$ in which the roles of
$a_1$ and of $a_2$ are interchanged. We may rewrite Eq.~(\ref{eq:A2rep3})
as
\begin{equation}
A_2^{c^2}(s,a_1,a_2) \sim
{\pi \over {\sqrt{a_1 a_2}}} \left[ {1 \over {s-1}} - \ln c^2 + \cdots \right]
+ {F_1}(a_1,a_2) \, , \qquad s \rightarrow 1 \,,
\end{equation}
where ${F_1}(a_1,a_2)$ is the finite, $s$-independent part of $A_2$ near
$s=1$, which
can be expressed as
\begin{equation}
{F_1}(a_1,a_2) = \frac{1}{{\sqrt{a_1 a_2}}}\left[f(a_1) + f(a_2) +
                    R(a_1,a_2) \right] \,,  \label{eq:F1def}
\end{equation}
where
\begin{equation}
f(a_1) = 4\pi \int_1^\infty {{(t^2-1)^{-\frac{1}{2}} dt}
       \over {{\rm e}^{2\pi c t/ \sqrt{a_1}} -1}} \,,    \label{frep}
\end{equation}
and
\begin{equation}
R(a_1,a_2) = 2\pi \sqrt{a_1} \sum^\infty_{n=-\infty}
\frac{1}{\sqrt{a_1 n^2+ c^2}\left(e^{2\pi\sqrt{(a_1 n^2+ c^2)/a_2}}-1\right)}
-f(a_2) \,, \label{eq:Rrep1}
\end{equation}
or, equivalently,
\begin{equation}
R(a_1,a_2) = 2\pi \sqrt{a_2} \sum^\infty_{n=-\infty}
\frac{1}{\sqrt{a_2 n^2+ c^2}\left(e^{2\pi\sqrt{(a_2 n^2+ c^2)/a_1}}-1\right)}
-f(a_1) \,. \label{eq:Rrep2}
\end{equation}
Note that $R(a_1,a_2) \rightarrow 0$ as $a_1 \rightarrow 0$ or as
$a_2 \rightarrow 0$. (The sum in Eq.~(\ref{eq:Rrep1}) may replaced by an
integral when $a_1 \rightarrow 0$, and that in Eq.~(\ref{eq:Rrep2}) may
be so replaced when $a_2 \rightarrow 0$.)

\end {document}